\newcommand{\sand}[3]{\langle#1|#2|#3 \rangle}

\def\beq {\begin{equation}}
\def\eeq {\end{equation}}
\def\bea {\begin{eqnarray}}
\def\eea {\end{eqnarray}}

\def\H {\mathcal{H}_{\mbox{\footnotesize eff}}}

\def\g {\frac{G_{F}}{\sqrt{2}}}

\def \Dppp {D^+ \to \pi^+ \pi^- \pi^+}

\documentclass{ws-ijmpe}

\begin{document}

\markboth{D. R. Boito, B. El-Bennich, B. Loiseau, O.
Leitner.}{Resonances and weak interactions in $D^+ \to\pi^+ \pi^- \pi^+$ decays.}

\catchline{}{}{}{}{}

\title{RESONANCES AND WEAK INTERACTIONS IN $D^+ \to\pi^+ \pi^- \pi^+$ DECAYS\footnote{Contribution to the X Hadron Physics, Florian\'opolis - Brazil, March 2007.} }

\author{\footnotesize DIOGO R. BOITO$^1$, BRUNO  EL-BENNICH$^2$, BENO\^IT LOISEAU$^2$, OLIVIER LEITNER$^{2,3}$. }

\address{$^1$Instituto de F\'isica, Universidade de S\~ao Paulo, C.P. 66318,\\
S\~ao Paulo, SP 05315-970, Brazil.\\
dboito@if.usp.br}

\address{$^2$ LPNHE-Laboratoire de Physique Nucl\'eaire et de Hautes \'Energies, Groupe Th\'eorie\\
IN2P3-CNRS, Universit\'es P. $\&$ M. Curie et Denis Diderot, 4 Pl. Jussieu, 75252 Paris, France\\
loiseau@lpnhe.in2p3.fr, bruno.elbennich@lpnhe.in2p3.fr}

\address{$^3$
INFN, Laboratori Nazionali di Frascati, \\
via E. Fermi, 40, 00044 Frascati (RM), Italy \\
olivier.leitner@lnf.infn.it}

\maketitle


\begin{abstract}
We describe the $\pi\pi$ $S$-wave in $\Dppp$ decays using a unitary
model for the $\pi\pi$ Final State Interactions (FSI). The three body
decay is treated as a {\it quasi two-body} process where, at the weak
vertex, the $D$ meson decays into a resonance and a pion. The weak
part of the decay amplitude is evaluated using the effective weak
Hamiltonian within the factorization approximation.
\end{abstract}

\section{Introduction}

The E791 collaboration found a strong evidence for a light
and broad scalar-isoscalar resonance in $D^+ \to \pi^+ \pi^-
\pi^+$ decays, known as the $\sigma(500)$ (or $f_0(600)$) meson~\cite{E791}. According to their fit, this resonance contributes to
approximately  half of the decays and has a mass and width given
by $m_\sigma = 478^{+24}_{-23} \pm 17 \,$ MeV and $\Gamma_\sigma =
328^{+42}_{-40}\pm 21\,$~MeV. This presence of the $\sigma(500)$
in $\Dppp$ has been confirmed by the CLEO collaboration in a very
recent paper~\cite{CLEO} and the results they obtained are in
agreement with those of E791.  On the other hand, this resonance was for a long
time hidden in scattering experiments, but at present its pole
position in the complex energy plane is rather well determined~\cite{Leutwyler}, 
which gives support to the E791 findings.

In our approach, we calculate the weak decay $D^+\to R\ \pi^+$
where $R=\sigma(500)$, $\rho(700)$, $f_0(980)$, using the
effective weak Hamiltonian within the
factorization approximation. The $(\pi\pi)_{S}$-wave  final state is then
constructed  using the scalar {\it form factor} previously introduced in
the context of $B$ decays to pions and kaons\cite{Furman,Bruno}.
This model for the FSI, based on the framework developed in~\cite{Meissner},
is unitary and takes into account the coupling to the $K
\bar K$ channel. The $P$-wave, which mainly corresponds  to the
$\rho(770)$, and the $D$-wave, arising from the $f_2(1270)$, are
described by the usual Breit-Wigner functions. The general idea of
the model is depicted in Fig. 1: the $\pi^+$ produced at the weak vertex acts as a 
spectator of the rescattering between the other two pions. The amplitude
is to be symmetrized since there are two identical pions in the
final state.
\begin{figure}[ht]
\centerline{\psfig{file=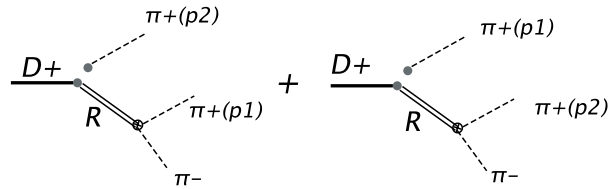,width=8cm}}
\vspace*{8pt}
\caption{$D^+$ decay to $\pi^+ \pi^- \pi^+$ intermediated by a
resonance $R$. The amplitude is symmetrized.}
\end{figure}

\vspace{-0.8cm}
\section{Weak amplitudes}

To evaluate the weak decays $D^+ \to (\sigma, \rho, f_0)\pi^+$, we use
an effective weak Hamiltonian within the QCD factorization
approximation. In the present case, the contributions from
strong penguins are CKM suppressed and can be safely neglected which  allows one
to work at tree level. Assuming that the $\sigma(500)$ is 
scalar-isoscalar with a $(u\bar u + d\bar d)$ quark content, the 
result for the decay $D^+ \to \sigma \pi^+$
can be cast as
\begin{equation}
\mathcal{A}_S= \sand{\sigma \pi^+}{\H}{D^+} = \g  V_{cd}V_{ud}^* \,a_1f_\pi(m^2_{D}
- m^2_{\sigma})F_0^{D\rightarrow \sigma}(m^2_\pi)\ ,
\label{eq:weakDsigma}
 \end{equation}
where $f_\pi$ is the pion decay constant, $m_D$, $m_\pi$ and
$m_\sigma$ are the $D$ meson, pion and  $\sigma$ masses, respectively,
$F_0^{D\rightarrow \sigma}$ is a transition form factor treated as a parameter of the model, $G_F$
is the Fermi constant and $V_{q_1q_2}$ is the CKM matrix element
for the transition $q_1\to q_2$.  The coefficient $a_1$ is a
combination of  Wilson coefficients, $C_1+ C_2/N_c$.  It is worth stressing that
this expression is simplified by the fact  it receives no contribution
from the color suppressed diagram since  the matrix element
$\sand{\sigma}{j^\mu}{0}$ with $j^\mu=\bar q \gamma^\mu(1-\gamma^5) q$ vanishes by symmetry. The result for the
weak decay $D^+\to f_0(980)\pi^+$ is very similar but one must also add the non-zero
$s\bar s$ component of the $f_0(980)$, which  can be done introducing a
mixing angle $\theta$ in the spirit of~\cite{Olivieretal}. The  $D^+ \to
\rho(770)^0\pi^+$ amplitude does receive a contribution from the color
suppressed diagram, proportional to  $a_2=C_2+ C_1/N_c$, since now
$\sand{\rho}{j^\mu}{0}\neq 0$.


\section{Final state interactions}

One can
construct the $3\pi$ final state using a form factor to describe
the propagation and decay into two pions of the resonance $R$. Usually a Breit-Wigner
function is used for this purpose \cite{Rogerio}. In this work, we employ a model 
for the $S$-wave in which four scalar
form factors are introduced  to describe the $(\pi\pi)_S$ and $(K \bar K)_S$ FSI~\cite{Furman,Bruno}. However, 
in the case being considered here, only one of them, namely
$\Gamma_1^{n*}(s)$, enters in the calculations. This model
respects unitarity and  the coupling to the $K
\bar K $ channel is properly accounted for. Employing an approximation where all the
particles are taken to be on-shell, this form factor is given analytically~\cite{Furman}: it is
written in terms of the scalar-isoscalar scattering phase shifts and inelasticities~\cite{Kaminski1997} $\delta_{\pi\pi}(s)$, $\delta_{K\bar
K}(s)$ and $\eta(s)$, as well as of
 the production functions $ R_1^{n}(s)$ and
$R_2^{n}(s)$ obtained in chiral perturbation theory
\cite{Meissner}. Below the $K \bar K$ threshold Watson's theorem
is fulfilled: the phase of the form factor $\Gamma_1^{n*}(s)$ is
given by $\delta_{\pi\pi}(s)$. The full amplitude $D^+ \to \sigma
\pi^+, \sigma \to \pi^+\pi^-$ can then be written as
\begin{equation}
 \mathcal{M}_S(u,t) = \mathcal{A}_S\ 
\chi \ \left [\ \Gamma_1^{n*}(u) + \Gamma_1^{n*}(t)\ \right ]\ ,
\label{eq:M}
\end{equation}
where $\mathcal{A}_S$ is given in Eq. (\ref{eq:weakDsigma}). This amplitude is symmetrized as shown in Fig.~1, it is expressed  in terms of  two Lorentz
invariant quantities namely $u = (p_1 + p_{\pi^-})^2$ and $t=(p_2+p_{\pi^-})^2$, following the labels used in Fig. 1. In Eq. (\ref{eq:M}), $\chi$ is a
normalization constant related to the coupling between the pions 
and the scalar resonance, estimated to be \cite{Furman} $\chi
\approx 30$ GeV$^{-1}$.
The amplitude for the decay $D^+\to \rho^0 \pi^+, \, \rho^0 \to
\pi^- \pi^+$ is written similarly
\begin{equation}
 \mathcal{M}_P(u,t) =  \mathcal{A}_P\,\left[\ (t-s)\Gamma_{\rho\pi\pi}(u)+ \, (u-s)
 \Gamma_{\rho\pi\pi}(t)\ \right ]\ ,
 \label{eq:Mrho}
\end{equation}
where $s=(p_1+p_2)^2$, $\Gamma_{\rho\pi\pi}$ is a relativistic Breit-Wigner
\cite{Bruno} and $\mathcal{A}_P$ the corresponding weak amplitude.  The $D$-wave is included in the model by means of
another Breit-Wigner function and two additional parameters: a magnitude
$a_{f_2}$ and a phase $\delta_{f_2}$.
The total amplitude is written as the sum of all the partial waves
\begin{equation}
\mathcal{M}_{\mbox{\tiny Total}}(u,t) =\mathcal{M}_S(u,t)+
\mathcal{M}_P(u,t)+ \mathcal{M}_D(u,t)\ . \label{eq:MTotal}
\end{equation}
This expression is in a form suited to obtain a Dalitz plot in terms of  $u$ and $t$.


\section{Preliminary results}

Eq. (\ref{eq:MTotal}) enables one to perform a fit to the E791 data
reproduced as in \cite{Oller}. We show in Tab. 1 preliminary results
for the normalization constant $\chi$ and for the transition form
factor $F_0^{D\to \sigma}$. The value for $\chi$ is in agreement with
the results from \cite{Furman,Bruno} whereas our value for $F_0^{D\to
\sigma}(m^2_\pi)$ is smaller than $0.79\pm0.15$ obtained in \cite{Rogerio}.

\begin{table}[ht]
\tbl{Preliminary results for $F_0^{D\to \sigma}(m_\pi^2)$ and for the
normalization constant $\chi$.}
{\begin{tabular}{@{}cc@{}} \toprule
Param. & Result  \\
\colrule
$F_0^{D\to\sigma}(m_\pi^2)$       & $0.43\pm 0.03 $ \\
$\chi $        &  $22.5\pm 0.8 $ GeV$^{-1}$    \\
\botrule
\end{tabular}}
\end{table}

In Tab.  2 we show our results for the fit fractions (f.f.s.), which
measure the contribution of each channel to the total Dalitz plot. Due
to interference effects, the sum of the f.f.s. is not necessarily
100\%. In our model, the $(\pi \pi)_S$ fit fraction approximatively
corresponds to the sum of the f.f.s. of $\sigma$, NR and $f_0's$ of
CLEO~\cite{CLEO}. Compared to Oller's work, we do not introduce a
non-resonant background. The description of the $S$-wave, unified as it
is in the $\Gamma_1^{n*}(s)$ form factor, avoids the use of a sum of
Breit-Wigner functions.  Use of a unitary model for the $\pi \pi$ FSI
enables us to reproduce the $S$-wave two-pion resonance spectrum in
$D^+\to \pi^+ \pi^- \pi^+$.
\vspace{-0.5cm}
\begin{table}[ht]
\tbl{Preliminary fit fractions  from this work compared with
E791, CLEO (isobar model fit)  and J. A. Oller  results. NR stands for non resonant.}
{\begin{tabular}{@{}ccccc@{}} \toprule
{\bf Channel} & {\bf E791 \cite{E791}(\%)}  & {\bf CLEO\cite{CLEO} (\%)}
 & {\bf Oller \cite{Oller}(\%)}  & {\bf This work(\%)}\\
\colrule
$\sigma$      & 46.3$\pm$ 9.2  & 41.8$\pm$ 1.4 $\pm$ 2.5  &      - & - \\
NR            & 7.8 $\pm$ 6.6  & $<$3.5 &   17 & -   \\
$f_0(980)$    & 6.2$\pm$ 1.4  &4.1$\pm$ 0.9 $\pm$ 0.3    &      - & - \\
$f_0(1370)$   & 2.3 $\pm$1.7 & 2.6 $\pm$1.8 $\pm$0.6   &   -    & - \\
$f_0(1500)$   & - &  3.4 $\pm$1.0  $\pm$ 0.8  &   -    & - \\
$(\pi\pi)_S$  & - &  -  &  102 & 53  \\
$\rho(770)$   & 33.6$\pm$3.9  & 20.0$\pm$2.3 $\pm$0.9     &   36 & 32 \\
$f_2(1270)$    & 19.4$\pm$2.5  & 18.2$\pm$2.6 $\pm$ 0.7    &   21 & 10 \\
$\rho(1440)$  & 0.7$\pm$0.8  &  $<$2.4  &    1 &  2\\
\botrule
\end{tabular}}
\end{table}

\vspace{-0.5cm}

\section*{Acknowledgments}

DRB thanks the LPNHE for the kind hospitality during the commencement  of this work and Dr. M. R. Robilotta
for discussions. DRB's work is supported by FAPESP (Brazilian Agency).

\end{document}